\documentclass[twocolumn,showpacs,preprintnumbers,amsmath,amssymb,floatfix]{revtex4}
\usepackage{graphicx}% Include figure files
\usepackage{dcolumn}% Align table columns on decimal point
\usepackage{bm}% bold math
\usepackage{color}

 \def\add#1{{{}{#1}}}                     % addition
\begin{document}

%\preprint{APS/123-QED}
%%%%%%%%%%%%%%%%%%%%%%%%%%%%%%%%%%%%%%%%%%%%%%%%%%%%%%%%%%%%%%%%%%
%%%%%%%%%%%%%%%%%%%%%%%%%%%%%%%%%%%%%%%%%%%%%%%%%%%%%%%%%%%%%%%%%%
\title{Shell to shell energy transfer in MHD,\\ 
Part I: steady state turbulence}% Force line breaks with \\
%%%%%%%%%%%%%%%%%%%%%%%%%%%%%%%%%%%%%%%%%%%%%%%%%%%%%%%%%%%%%%%%%%
%%%%%%%%%%%%%%%%%%%%%%%%%%%%%%%%%%%%%%%%%%%%%%%%%%%%%%%%%%%%%%%%%%
\author{Alexandros Alexakis}
\email{alexakis@ucar.edu}
\author{Pablo D. Mininni}%
 \email{mininni@ucar.edu}
\author{Annick Pouquet}
             \email{pouquet@ucar.edu}
\affiliation{National Center for Atmospheric Research, 
             P.O. Box 3000, Boulder, Colorado 80307}
%%%%%%%%%%%%%%%%%%%%%%%%%%%%%%%%%%%%%%%%%%%%%%%%%%%%%%%%%%%%%%%%%%
%%%%%%%%%%%%%%%%%%%%%%%%%%%%%%%%%%%%%%%%%%%%%%%%%%%%%%%%%%%%%%%%%%
\date{\today}
%%%%%%%%%%%%%%%%%%%%%%%%%%%%%%%%%%%%%%%%%%%%%%%%%%%%%%%%%%%%%%%%%%

\begin{abstract}
%\opt{The assumption of locality of transfer of energy among the different 
%scales involved in a turbulent flow is one of the building blocks of 
%Kolmogorov (1941) theory of turbulence . This assumption for 
%hydrodynamic turbulence has been carried over to present models of 
%magneto-hydrodynamical turbulence. However, if this assumption is true 
%for conducting fluids in the presence of magnetic fields has to be
%tested.} 
We investigate the transfer of energy from large 
scales to small scales in
fully developed forced three-dimensional MHD-turbulence
by analyzing the results of direct numerical simulations
 in the absence of an externally imposed 
uniform magnetic field. Our results show that the transfer of 
kinetic energy from the large scales to kinetic energy at smaller scales, 
and the transfer of magnetic energy from the large scales to magnetic 
energy at smaller scales, are local, as is also found in the case of neutral fluids, 
and in a way that is compatible with Kolmogorov (1941) theory of turbulence. 
However, the transfer of energy from 
the velocity field to the magnetic field is a highly non-local process in Fourier space.
Energy from the velocity field at large scales can be transfered directly
into small scale magnetic fields without the participation of intermediate 
scales. Some implications of our 
results to MHD turbulence modeling  are also discussed.
\end{abstract}

% PACS, the Physics and Astronomy
                             % Classification Scheme.
%\keywords{Suggested keywords}%Use showkeys class option if keyword
                              %display desired
\maketitle

%%%%%%%%%%%%%%%%%%%%%%%%%%%%%%%%%%%%%%%%%%%%%%%%%%%%%%%%%%%%%%%%%%%%%%%%%%%%%%%%%%%%%%
%%%%%%%%%%%%%%%%%%%%%%%%%%%%%%%%%%%%%%%%%%%%%%%%%%%%%%%%%%%%%%%%%%%%%%%%%%%%%%%%%%%%%%
\section{ \label{Intro} Introduction }

Most astrophysical and planetary systems, e.g. solar/stellar winds, 
accretion disks  and interstellar medium, are in a turbulent state and
coupled to magnetic fields. Understanding and quantifying
the statistical properties of magnetohydrodynamical (MHD) turbulence
is crucial to explain many physical processes 
in the cosmos, and in industrial flows as well \cite{Davidson}. 
Although the
phenomenology of hydrodynamical (HD) turbulence is understood to some 
extent, and the theory has been able to make predictions 
like Kolmogorov's 4/5th law and the functional form of the energy spectrum
in the inertial range,
that have been well verified in experiments and numerical simulations, a similar 
statement cannot be made for MHD turbulence at the same level. In MHD 
flows, the two fields (velocity and magnetic) and two associated energies involved in the 
dynamical processes allow for many possibilities for the energy to transfer between
smaller or larger scales, making the dynamics more complex 
to address in both theory and modeling.

We briefly describe some phenomenological aspects of HD turbulence to 
point out some of the difficulties usually encountered when the 
formulation of HD turbulence is applied in the MHD case. To follow 
Kolmogorov (1941)  theory \cite{K41} (hereafter, K41), we need to assume 
a statistically isotropic and homogeneous flow in steady state in which 
the energy is cascading from eddies of scale $l$ to smaller eddies, and 
so on until energy reaches the dissipation scales. Since we are 
considering a statistically steady state, 
the flux of energy to smaller scales has to be constant. We can further assume 
that the flux at some scale can depend only on the scale $l$ and the 
amplitude of the velocity field $|{\bf u}_l|$ at this scale. This assumption 
is justified by the argument that larger eddies will only advect smaller 
eddies without significantly altering their scale 
and only when eddies of similar size interact do they produce a cascade. 
Therefore, only 
``local" interactions among the different scales control the cascade. 
Here we use the term ``local" in terms of the different scales 
involved (i.e. scales of similar size) and not as locality in physical space.
With these assumptions we obtain that the energy $|{\bf u}_l|^2$ at the 
scale $l$ will cascade to smaller scales in a time $l/|{\bf u}_l|$, and 
since the energy cascade rate $\epsilon$ is constant, we obtain 
$\epsilon \sim |{\bf u}_l|^3/l$ that implies $|{\bf u}_l|\sim l^{1/3}$, which 
finally leads to the well verified K41 spectrum $dE/dk \sim k^{-5/3}$ 
to within small intermittency corrections. 

The assumptions of the HD theory of turbulence have been tested in the 
literature. Ref. \cite{Domaradzki88} first tested the assumption of 
locality using direct numerical simulations (DNS) of $64^3$ grid points.
Their work has been followed by a number of authors with higher resolution 
simulations 
\cite{Domaradzki90,Ohkitani92,Zhou93L,Zhou93,Yeung95,Zhou96,Kishida99}.
Refs. \cite{Yeung91,Brasseur94,Yeung95} have also investigated the effect of 
long-range interactions and anisotropy induced by an anisotropic large 
scale flow. Although some issues still remain regarding the effect of 
long range interactions, the locality of the energy transfer has been 
confirmed.

However, there are two important assumptions used in the HD
case that are not necessarily true for the MHD case.
First, the assumption of isotropy breaks down if an imposed uniform  magnetic
field is considered. We will not investigate such effects in
the present work and will only consider flows with 
$\int {\bf b} \, d{\bf x}^3=0$.
The second assumption, that of   locality
of interactions among the different scales is what  motivates our work. Unlike the HD case where 
the effect of larger eddies on smaller ones is the advection of the later 
ones (an effect that can be taken away by a Galilean transformation), in 
MHD the effect of a large scale fluctuation of the magnetic field cannot be so 
eliminated. Therefore, in MHD it is possible for small scales to 
interact directly with the large scales. If this is the case, we can not consider 
a ``contiguous" transfer of energy in wave number space and cannot {\it a priori} follow
the same arguments Kolmogorov used for HD turbulence.
Therefore knowledge of the energy transfer among different scales 
is important for the construction of any phenomenological model of 
turbulence.

Present phenomenological models follow Kolmogorov like arguments
that take into account the effect of the magnetic field.
Iroshnikov \cite{Iroshnikov}
and Kraichnan \cite{Kraichnan} proposed the first models to describe
isotropic-MHD turbulence, predicting a spectrum of $k^{-3/2}$ 
(hereafter, IK). 
Goldreich and Shridar \cite{Goldreich} proposed a new model
for anisotropic MHD turbulence that takes into account the anisotropy
introduced by a uniform magnetic field ${\bf B}_0$, predicting a spectrum of
$k_\perp^{-5/3}$, where $k_{\perp}$ refers to the direction perpendicular to ${\bf B}_0$. 
Several models have been proposed that combine the
two spectra (see e.g. \cite{Matthaeus,Boldyrev,Galtier05}), suggesting that the
index of the energy spectrum is sensitive to the presence and intensity of ${\bf B}_0$.
Some aspects of non-locality of interactions are taken into account 
in the afore-mentioned models by considering that large scale fluctuations of 
the magnetic field act as a uniform magnetic field to the smaller scales,
and as a result they speed up or slow down the rate at which the energy is 
cascading. However, in these models, although non-local interactions
are taken into account, the energy is transfered 
locally from one scale to a slightly smaller scale, like in Kolmogorov's
HD turbulence model. 

The locality of the interactions and the energy transfer in MHD turbulence
has been investigated through various closure models. The energy transfer
has been studied within the EDQNM closure model by \cite{Pouquet}
and more recently by \cite{Schilling} where non-local interactions have been noted. 
Using field theoretical calculations the transfer of energy has been estimated by
\cite{Verma01,Verma03,Verma05}. 
As far as we know, the locality of the energy transfer in MHD
has been investigated through three dimensional direct numerical simulations (DNS)  only very 
recently \cite{Debliquy} (see also \cite{Dar01} for the two-dimensional case).
These authors measured the transfer of energy between different scales and fields 
using free decaying MHD turbulence simulations with $512^3$ grid points.
Their results showed that there is local transfer of energy between 
the same fields, while the transfers involving the two different fields 
showed a less local behavior, in the sense that a wider range of scales 
was involved in the interactions.

In our work we use the results of DNS of mechanically forced MHD turbulence 
(\add{unlike the free decaying case studied in \cite{Verma05}}) 
to study the locality of the energy transfer between different scales 
and fields.
In all the cases studied we consider a mechanic external forcing that
generates a well defined large scale flow and small scale turbulent
fluctuations.  This is a regime of interest for several astrophysical and geophysical flows 
where
magnetic fields are believed to be sustained against Ohmic dissipation by a dynamo process 
\cite{Moffatt}, and the only external source of energy driving the system 
is mechanical (e.g. convection and rotation).
There is an important difference between the case studied in \cite{Debliquy} 
and the case considered in our work. In our case energy is forced through 
the velocity field and the system reaches a steady state with equipartition 
between the two fields. For this to happen there must be a non-zero 
flux for all times from the velocity field
to the magnetic field. This is not necessarily true for the case of decaying turbulence 
and as our results show this significantly modifies the energy transfers from the velocity 
field to the magnetic field.

In Sec. \ref{Theory} 
we introduce the definitions of the transfer terms for MHD, and in 
Sec. \ref{Results} we present the code we use for the numerical 
simulations as well as the results of the analysis. Finally, in 
Sec. \ref{Concs} we summarize the main results of our work.

\section{ \label{Theory} Theory and definitions}

The equations that describe the dynamics of an 
incompressible conducting fluid coupled to a magnetic field in the MHD approximation are given by:
\begin{equation}
\partial_t {\bf u} + {\bf u}\cdot \nabla {\bf u} = - \nabla p + 
{\bf b}\cdot \nabla {\bf b} + \nu \nabla^2 {\bf u} +{\bf f} ,
\label{eq:momentum}
\end{equation}
\begin{equation}
\partial_t {\bf b} + {\bf u}\cdot \nabla {\bf b} = {\bf b}\cdot \nabla {\bf u}
+ \eta \nabla^2 {\bf b} ,
\label{eq:Induction}
\end{equation}
\begin{equation}
\nabla \cdot {\bf u} =0, \,\,\, \nabla \cdot {\bf b} =0 ,
\label{eq:Incompresible}
\end{equation}
where ${\bf u}$ is the velocity field and ${\bf b}$ is the magnetic field. 
$p$ is the (total) pressure and $\nu$ and $\eta$ are the viscosity and the 
magnetic diffusivity respectively. 
Here, ${\bf f}$ is the external force that drives the turbulence and the 
dynamo. 
The largest wavenumber of the Fourier transform of $f$ is going to be denoted
as ${\bf k}_F$ and we are going to refer to ${|\bf k}_F|^{-1}$ as the forced scale.
We are also going to define the viscous dissipation scale as  
$k_\nu^{-1}=(\epsilon/\nu^3)^{-1/4}$ and resistive dissipation scale as 
$k_\eta^{-1}=(\epsilon/\eta^3)^{-1/4}$ where $\epsilon$ is the energy dissipation rate. 
A large separation between the two 
scales (${|\bf k}_F|^{-1} \gg \max \{ \{k_\nu^{-1},k_\eta^{-1} \} $) is required 
for the flow to reach a turbulent state.

To investigate the transfer of energy among different
scales of turbulence we use the Fourier transforms
of the fields: 
\[ {\bf       {u(x)}} = \sum_{\bf k} {\bf \tilde{u}(k)}e^{i{\bf kx}} 
\,\,\, {\mathrm{,} } \,\,\,
{\bf \tilde{u}(k)} = \frac{1}{({2\pi})^3}\int {\bf {u}(x)}e^{-{\bf ikx}} 
d{\bf x}^3 \,\,\, \]
and 
\[ {\bf       {b(x)}} = \sum_{\bf k} {\bf \tilde{b}(k)}e^{i{\bf kx}} 
\,\,\, {\mathrm{,} } \,\,\,
{\bf \tilde{b}(k)} = \frac{1}{({2\pi})^3}\int {\bf {b}(x)}e^{-i{\bf kx}} 
d{\bf x}^3 \,\,\, , \]
where the domain is taken to be a triply periodic cube of size $L=2\pi$.
We can now introduce the shell filter decomposition:
\[ {\bf u(x)} = \sum_K {\bf u}_K(x) , \,\, {\bf b(x)} = \sum_K {\bf b}_K(x) \]
where 
\[{\bf u}_K(x)= \sum_{K<|{\bf k}|\le K+1} {\bf \tilde{u}(k)}e^{i{\bf kx}} , \]
and similar for the field ${\bf b}$ 
\[{\bf b}_K(x)= \sum_{K<|{\bf k}|\le K+1} {\bf \tilde{b}(k)}e^{i{\bf kx}} . \]
The fields ${\bf u}_K$ and ${\bf b}_K$ are therefore defined as the part 
of the velocity 
and magnetic field respectively, whose Fourier transform contains only 
wave numbers in the shell $(K, K+1]$ (hereafter called the shell $K$)
and represent ``eddies" of scale $K^{-1}$. The evolution of the kinetic energy in a shell $K$, 
$E_u(K)=\int {\bf u}_K^2/2 \, dx^3$ is given by:
\begin{eqnarray}
\partial_t E_u(K) &=& 
   \int \sum_Q \left[- {\bf u}_K {\bf  \cdot (u \cdot \nabla) \cdot u}_Q
                     + {\bf u}_K {\bf  \cdot (b \cdot \nabla) \cdot b}_Q \right]
 \nonumber \\
{ } &&              - \nu |\nabla {\bf u}_K|^2
                    + {\bf f \cdot u}_K \,\, d{\bf x}^3  \, ,
\end{eqnarray}
and for the magnetic energy  $E_b(K)=\int {\bf b}_K^2/2 \, dx^3$ 
we obtain
\begin{eqnarray}
\partial_t E_b(K) &=& 
   \int \sum_Q \left[- {\bf b}_K {\bf  \cdot (u \cdot \nabla) \cdot b}_Q
                     + {\bf b}_K {\bf  \cdot (b \cdot \nabla) \cdot u}_Q\right]
    \nonumber \\
{ } &&              - \eta |\nabla {\bf b_K}|^2 \,\, d{\bf x}^3 \, .
\end{eqnarray}
The above equations can be written in the more compact form:
\begin{equation}
\partial_t { E}_{u}(K) = 
\sum_Q [{\mathcal T}_{uu}(Q,K)+{\mathcal T}_{bu}(Q,K)] - \nu {\mathcal D}_u(K) 
+ {\mathcal F}(K) ,
\label{eq:Eu}
\end{equation}
\begin{equation}
\partial_t { E}_{b}(K) = 
\sum_Q [{\mathcal T}_{ub}(Q,K)+{\mathcal T}_{bb}(Q,K)] - 
\eta {\mathcal D}_b(K).
\label{eq:Eb}
\end{equation}
Here we have introduced the functions ${\mathcal T}_{uu}(Q,K)$, 
${\mathcal T}_{ub}(Q,K)$, ${\mathcal T}_{bb}(Q,K)$, and 
${\mathcal T}_{bu}(Q,K)$ 
that express the energy transfer between different fields and shells.

${\mathcal T}_{uu}(Q,K)$ expresses the transfer rate of kinetic energy 
lying in the shell $Q$ to kinetic energy lying in the shell $K$
through the velocity advection term and is defined as:
\begin{equation}
{\mathcal T}_{uu}(Q,K) \equiv 
-\int {\bf {u}_K (u \cdot \nabla) {u}_Q  } d{\bf x}^3 .
\label{eq:Tuu}
\end{equation}
We similarly define
\begin{equation}
{\mathcal T}_{bb}(Q,K) \equiv
-\int {\bf {b}_K (u \cdot \nabla) {b}_Q  } d{\bf x}^3, 
\label{eq:Tbb}
\end{equation}
which expresses the rate of energy transfer of magnetic energy lying in 
the shell Q to magnetic energy lying in the shell $K$ through the 
magnetic advection term. 
The Lorentz force is responsible for the transfer 
of energy from the magnetic field to the velocity field. The resulting 
transfer rate is defined as: 
\begin{equation}
{\mathcal T}_{bu}(Q,K) \equiv
\int {\bf {u}_K (b \cdot \nabla) {b}_Q  } d{\bf x}^3.
\label{eq:Tub}
\end{equation}
Finally the term responsible for the stretching of the magnetic field lines 
results in the transfer from kinetic energy to magnetic energy, given by:
\begin{equation}
{\mathcal T}_{ub}(Q,K) \equiv
\int {\bf {b}_K (b \cdot \nabla) {u}_Q  } d{\bf x}^3 .
\label{eq:Tbu}
\end{equation}
In summary,  the functions ${\mathcal T}_{vw}(Q,K)$ (for arbitrary fields 
${\bf v}$ and ${\bf w}$) represent the rate of  transfer of energy
from the field ${\bf v}$ (first index) in the shell $Q$ (first argument), 
into energy of the field ${\bf w}$ (second index) in the shell $K$ (second argument).
If ${\mathcal T}_{vw}(Q,K)>0$, then a positive amount of $v-$energy is transfered 
from the shell $Q$ to $w-$energy in the shell $K$. 
If ${\mathcal T}_{vw}(Q,K) < 0$, then 
a negative amount of $v-$energy is transfered from the shell $Q$ to $w-$energy in the shell 
$K$, or in other words, energy is transfered backwards from the shell $K$ to the shell $Q$.

In eqs. (\ref{eq:Eu}-\ref{eq:Eb}) we have also introduced two
dissipation functions: the kinetic energy dissipation rate
\begin{equation}
\nu {\mathcal D}_{u}(K) \equiv  \nu   \int |{\bf \nabla u}_K|^2  d{\bf x}^3 ,
\end{equation}
and the magnetic energy dissipation rate
\begin{equation}
\eta {\mathcal D}_{b}(K) \equiv  \eta \int |{\bf \nabla b}_K|^2 d{\bf x}^3 .
\end{equation}
Finally, 
\begin{equation}
{\mathcal F}(K) \equiv \int { \bf f} \cdot {\bf u}_K  \, d{\bf x}^3
\end{equation}
is the energy injection rate to the velocity field through the forcing term.

Before presenting the results from numerical simulations, let us discuss 
some of the properties of the transfer functions. If ${\mathcal T}_{vw}(Q,K)$
(where $v,w$ can be either $u$ or $b$) is expressing the rate of energy 
transfer from the field ${\bf v}$ in the shell $Q$ to the field ${\bf w}$ 
in the shell $K$, then the following identity should hold  
\begin{equation}
\label{tran_id}
{\mathcal T}_{vw}(Q,K)=-{\mathcal T}_{wv}(K,Q) .
\end{equation}
The interpretation of eq.(\ref{tran_id}) is
that the rate at which the shell $Q$ is giving energy to the shell 
$K$ must be equal to the rate the shell $K$ is receiving energy from the 
shell $Q$. Eq. (\ref{tran_id}) can be easily shown to hold for all the 
transfer functions we defined (eqs. [\ref{eq:Tuu}-\ref{eq:Tbu}]). It is 
this property that allows us to interpret the functions ${\mathcal T}_{uu}$, 
${\mathcal T}_{bu}$, ${\mathcal T}_{ub}$, and ${\mathcal T}_{bb}$ as the 
energy transfer between different scales and fields. 

For a turbulent flow in a statistically steady state, 
equations (\ref{eq:Eu}) and (\ref{eq:Eb}) imply that:
\begin{equation} 
\sum_Q\langle{\mathcal T}_{uu}(Q,K)+{\mathcal T}_{bu}(Q,K) \rangle =
\langle {\mathcal D}_u(K) \rangle - \langle {\mathcal F}(K)\rangle ,
\end{equation}
and
\begin{equation} 
\sum_Q\langle{\mathcal T}_{ub}(Q,K)+{\mathcal T}_{bb}(Q,K) \rangle=
\langle {\mathcal D}_b(K) \rangle ,
\end{equation}
where $\langle \cdot \rangle$ stands for a time average or an ensemble 
average. For fixed $K$ outside the forcing band, and in the limit of 
$\nu,\eta \to 0$, we have that
\begin{equation} 
\sum_Q\langle{\mathcal T}_{uu}(Q,K)+{\mathcal T}_{bu}(Q,K) \rangle = 0
\end{equation}
and
\begin{equation} 
\sum_Q\langle{\mathcal T}_{ub}(Q,K)+{\mathcal T}_{bb}(Q,K) \rangle=  0 .
\end{equation} 
However, limited resolution will allow us to be in the regime
where these last two equations hold only for a small range of wavenumbers.

Finally we need to comment on the definitions of the various transfer 
functions we are using in this paper and the connection to the triad of
wave numbers (${\bf k,p,q}$) that satisfy the relation ${\bf k+p+q=0}$ 
(because of the convolution term resulting from the quadratic nonlinearities 
of the primitive equations); such triad is the basis for mode to mode interactions 
(see e.g. \cite{Kraichnan2}). 
Our approach is equivalent to 
considering all triad interactions with the one wavenumber ${\bf k}\in K$
and ${\bf q} \in Q$ and summing over all ${\bf p}$ satisfying 
${\bf k+p+q=0}$ in all shells, where ${\bf p}$ is the wave-number of 
the advecting field, and ${\bf k}$ and ${\bf q}$ 
are the wavenumbers of the modes energy is transfered to and from.
Although the approach we are using gives us information 
on whether the energy is transfered locally or not, 
it cannot give definite conclusions on whether the interactions themselves
are local. For example, even if energy is transfered locally from a wavenumber
${\bf k}$ to a wavenumber ${\bf q\sim k}$,
the wavenumber ${\bf p}$ that is responsible for the transfer
is not necessarily of the same order of magnitude as $|{\bf k}|$ and $|{\bf q}|$.
Ideally, one would investigate transfer terms of the form: 
${\bf T}_{uu}(K|P|Q)\equiv \int {\bf u}_K({\bf u}_P \cdot \nabla) {\bf u}_Q d{\bf x}^3$ 
that contain information about the third wave number involved in the 
interactions taking place. However the difficulty of manipulating data 
from high resolution runs and the difficulty of interpreting the results 
of transfer functions that depend on three arguments restricts us for the present time, to 
examine just the locality of the energy transfer. 

\section{ \label{Results} Results}

To study the transfer of energy in MHD turbulence we use the 
turbulent steady state of several mechanically forced three dimensional 
MHD direct numerical simulations. The simulations and details of the 
code can be found in \cite{MininniApJ,MininniPRE}. The runs were 
performed in a triply periodic domain with a resolution of $256^3$ grid 
points, using a pseudo-spectral scheme with the $2/3$-rule for dealiasing. 
The equations were evolved in time using a second order Runge-Kutta method.

Turbulence was generated by two different types of forcing. In the first 
case a non-helical Taylor-Green force (hereafter referred as TG) was used 
${\bf f}_{\rm TG}(k_0)=(\, \sin(k_0 x) \cos(k_0 y) \cos(k_0 z), 
                          -\cos(k_0 x) \sin(k_0 y) \cos(k_0 z),
                            0)$ 
with $k_0=2$
\cite{MininniApJ}. In the second case a helical ABC force was used 
${\bf f}_{ABC}(k_0) = (\,B\cos(k_0 y)+C\sin(k_0 z),\,
                         C\cos(k_0 z)+A\sin(k_0 x),\, 
                         A\cos(k_0 x)+B\sin(k_0 y)\,)$
with $k_0=2$
\cite{MininniPRE}. All simulations were done with constant in time external 
force. First a hydrodynamic simulation was carried using each force, to reach 
a turbulent steady state. Both external forces generate a well defined 
large scale flow at $|K_F|\sim3$, and small scale turbulent fluctuations 
following to a good approximation a 5/3 Kolmogorov law. 
Then MHD simulations were carried, and a 
small magnetic field was amplified and sustained to equipartition by a 
dynamo process. The results in this paper are based on the saturated 
stage of the dynamo, which we will refer in the following as the MHD 
turbulent steady state.
 
\begin{figure}
\includegraphics[width=8cm]{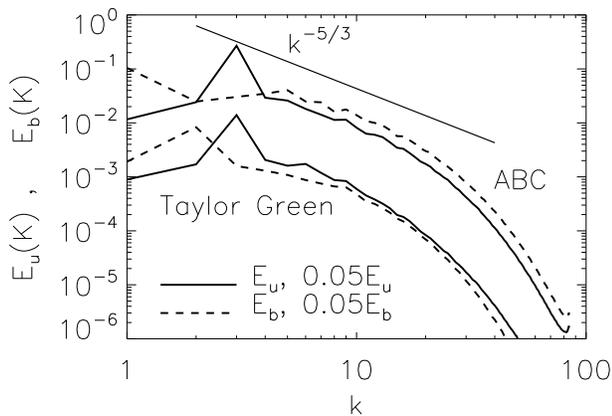}
\caption{\label{fig_01}Spectra of kinetic energy (solid line) 
and magnetic energy (dashed line) of the ABC and Taylor Green runs,
where the Taylor Green spectra have been shifted down by a factor of 20
for clarity. The Kolmogorov slope is showed as a reference. 
Note that the magnetic Prandtl number $P_M \equiv \nu/\eta$ differs for the two runs.}
\end{figure}

The transfers were calculated based on the definitions 
(\ref{eq:Tuu}--\ref{eq:Tbu}).
The transfer of energy during the early 
stages of the MHD simulations, when the magnetic energy is small and the 
velocity field is not modified by the Lorentz force (often referred to as 
the kinematic dynamo regime) are examined in a companion paper \cite{us} 
(hereafter referred as Paper II). Table \ref{runs} gives several relevant parameters for 
each run, and figure \ref{fig_01} shows the resulting energy spectra.

Both simulations display a large scale magnetic field, although the 
spectrum of magnetic energy in the ABC simulation shows a stronger peak 
at $k=1$. This peak is related with the dynamo $\alpha$-effect 
and the inverse cascade of magnetic helicity.
 Details of this process will be discussed 
in Paper II. However, it is important to note that in the ABC simulation 
the large scale magnetic field is strongly helical, while in the TG 
simulation the magnetic helicity is negligible. This large scale magnetic 
field is self-sustained by the turbulence. In both simulations, the 
net cross helicity (correlation between the velocity and the magnetic field) 
is small and can be neglected.

\begin{table}
\caption{\label{runs}Simulations. $L$ is the integral length-scale of the 
flow, defined as $L=\int{E_u(k)\, dk}/\int{E_u(k) k^{-1} \, dk}$; $\nu$ 
is the kinematic viscosity, and $\eta$ the magnetic diffusivity. 
The kinetic and magnetic Reynolds numbers $Re$ and R$_M$ are based on $L$ 
and the rms velocity, while the ratio of magnetic to kinetic energy 
$E_b /E_u$ is the average in the turbulent steady state.}
\begin{ruledtabular}
\begin{tabular}{lcccccc}
Forcing & $L$  &      $\nu$      &     $\eta$       & Re  & R$_M$&$E_b/E_u$\\
\hline
ABC     &$1.64$&$2\times10^{-3}$ & $2\times10^{-3}$ & 820 & 820  & $0.84$  \\
TG      &$1.35$&$2\times10^{-3}$ & $5\times10^{-3}$ & 675 & 270  & $0.72$  \\
\end{tabular}
\end{ruledtabular}
\end{table}
%%%%%%%%%%%%%%%%%%%%%%
%%%%%%%%FIGURE 1%%%%%%
%%%%%%%%%%%%%%%%%%%%%%

%%%%%%%%%%%%%%%%%%%%%%%%%%%%%%%%%%%%%%%%%%%%%%%%%%%%%%%%%%%%%%%%%%%
%%%%%%%%%%%%%%%%%%%%%%%%%%%%%%%%%%%%%%%%%%%%%%%%%%%%%%%%%%%%%%%%%%%
\subsection{ \label{Hydro} Hydrodynamic Turbulence } 

Locality of interactions in hydrodynamic turbulence have been investigated
before in the literature 
\cite{Domaradzki90,Ohkitani92,Zhou93L,Zhou93,Yeung95,Zhou96,Kishida99}. 
Although some open issues still remain 
%about the persistence of some long range interactions 
\cite{Yeung91,Brasseur94,Yeung95} 
it has been shown that energy is transfered mostly locally. 
Here, for reasons of comparison we show the transfer
${\mathcal T}_{uu}(Q,K)$ from hydrodynamical simulations using the same 
external forces and parameters used in the MHD simulations. The results 
are in good agreement with previous works.

\begin{figure}
\includegraphics[width=8cm]{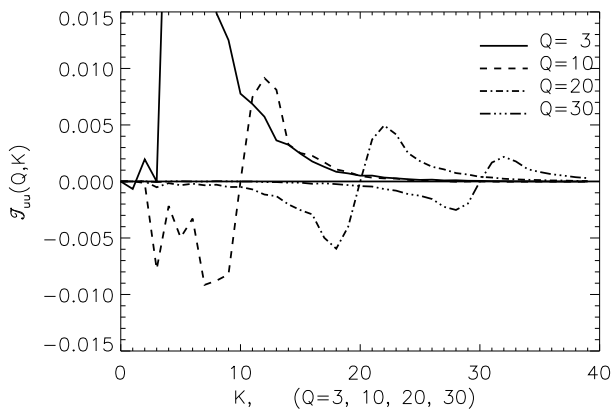}
\caption{\label{fig_02} The transfer of energy ${\mathcal T}_{uu}(Q,K)$
for the Taylor-Green run. The figure shows the rate that energy is
transfered from the modes $Q=3,10,20,30$  to all the other modes $K$.}
\end{figure}

\begin{figure}
\includegraphics[width=8cm]{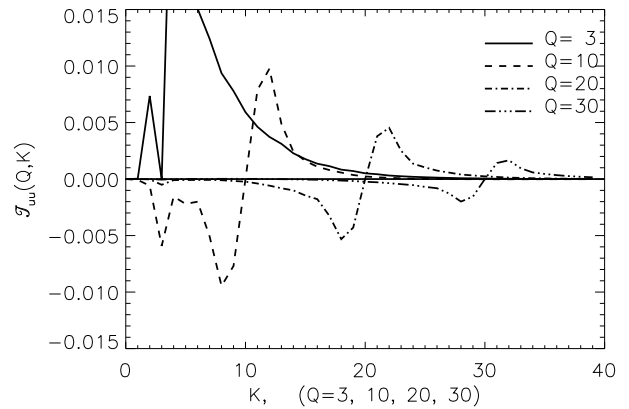}
\caption{\label{fig_03} The transfer of energy ${\mathcal T}_{uu}(Q,K)$
for the ABC run. The figure shows the rate that energy is
transfered from the modes $Q=3,10,20,30$  to all the other modes $K$.}
\end{figure}

In figure \ref{fig_02} we show the energy transfer for a few modes for 
the TG flow and in figure \ref{fig_03} the energy transfer 
for the ABC flow.
In both cases the transfer of energy is direct and local:
all the curves (with the exception of the forced mode $Q=3$) 
are negative for $K$ smaller than $Q$ 
and positive for $K$ larger than $Q$.
As a result, all inertial range modes 
receive energy from modes with slightly smaller wavenumbers 
(negative ${\mathcal T}_{uu}$) 
and give energy to modes with slightly 
larger wave number (positive ${\mathcal T}_{uu}$). 
The locality of 
the transfer is expressed from the fact that the transfer of 
energy from the modes in the shell $Q$ to modes in shells $K$ with 
$K \ll Q$ or $K \gg Q$ is very small, and decreases fast with the 
separation of the two wave numbers.
Finally, as the shell wavenumber $K$ and $Q$ is increased, there is
a drop in the amplitude of the transfer. If the transfer functions were
self-similar then an increase of the wave numbers $K$ and $Q$ to
$\lambda K$ and $\lambda Q$ would imply ${\mathcal T}_{uu}(\lambda Q,\lambda K)=
\lambda^{-2}{\mathcal T}_{uu}(Q , K)$ \cite{Kraichnan2}. 
This scaling could explain this drop of amplitude. However 
the inertial range in our DNS is too small to test self-similarity and 
a large part of the drop is due to the presence of viscosity.

The forced mode has a slightly different behavior. The transfer 
rate from the forced wave number to its nearby shells has a 
considerably larger amplitude. Also, for both flows there 
is some backscattering from the forced wave number to shells with 
smaller wavenumber. This is clearer in the helical (ABC) flow. 

\subsection{ \label{MHD} Magneto-Hydrodynamic Turbulence }

We are now ready to examine results from the energy transfer for MHD 
turbulence. First we examine the transfer of kinetic energy from large 
scales to kinetic energy in small scales through the term ${\mathcal T}_{uu}(Q,K)$, and 
magnetic energy from large scales to magnetic energy in small scales 
through the term ${\mathcal T}_{bb}(Q,K)$. These two transfer functions bare 
some significant similarities with the hydrodynamic case.

\begin{figure}
\includegraphics[width=8cm]{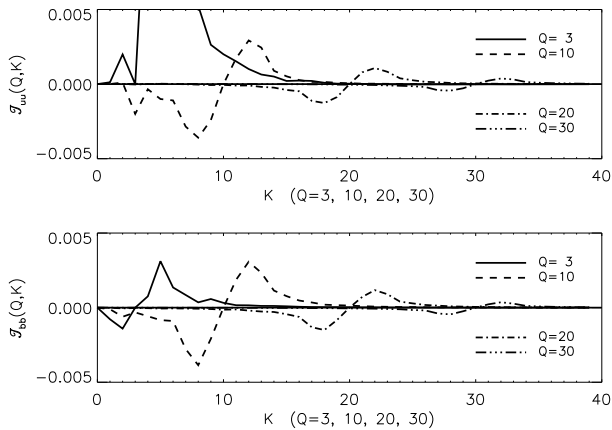}
\caption{\label{fig_04} Top panel:The transfer of energy 
${\mathcal T}_{uu}(Q,K)$ for the Taylor-Green run.The figure shows 
the rate that kinetic energy is transfered from the modes $Q=3,10,20,30$ 
to kinetic energy to all the other modes $K$. 
Bottom panel: The transfer of energy ${\mathcal T}_{bb}(Q,K)$ for the same flow. The 
figure shows the rate that magnetic energy is transfered from the 
modes $Q=3,10,20,30$ to magnetic energy to all the other modes $K$.}
\end{figure}

\begin{figure}
\includegraphics[width=8cm]{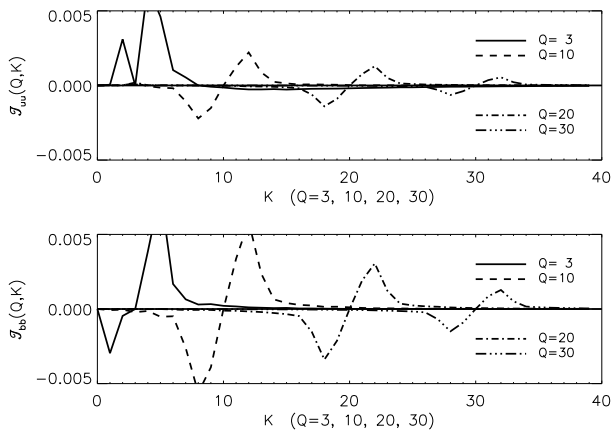}
\caption{\label{fig_05} Same as figure 4 for the ABC run} 
%\caption{\label{fig_05} Top panel:The transfer of energy 
%${\mathcal T}_{uu}(Q,K)$ for the ABC run.The figure shows the 
%rate that kinetic energy is transfered from the modes $Q=3,10,20,30$ 
%to kinetic energy to all the other modes $K$. 
%Bottom panel: The transfer of energy ${\mathcal T}_{bb}(Q,K)$. The 
%figure shows the rate that magnetic energy is transfered from the modes 
%$Q=3,10,20,30$ to magnetic energy to all the other modes $K$}
\end{figure}

In figures \ref{fig_04} and \ref{fig_05} we show ${\mathcal T}_{uu}$ 
(top panel) and ${\mathcal T}_{bb}$ (bottom panel) for the non-helical 
TG flow and the helical ABC flow.
%%%% FIGURE 4 %%%%%
%%%%%%%%%%%%%%%%%%%
%%%% FIGURE 5 %%%%%
%%%%%%%%%%%%%%%%%%%
The velocity to velocity transfer has not changed drastically
(other than a decrease in amplitude) from the pure hydrodynamic case. 
 As in Sec. \ref{Hydro}, the transfer implies a local direct cascade.
All the curves are negative for $K$ smaller than $Q$, 
               and positive for $K$ larger  than $Q$. 
Each mode is therefore receiving energy 
from the larger scales (negative transfer) and giving energy to the 
smaller scales (positive transfer). The decrease in amplitude 
(when compared with the 
hydrodynamic case) is partly because the magnitude of the velocity 
field is decreased when magnetic field comes to equipartition,
% with the velocity field, 
and partly because now there is a net transfer 
of energy from the velocity field to the magnetic field, making the 
available energy to cascade to small velocity scales smaller.
%Note also a small backscattering from the small scales to
%$K$-shells larger than the forcing band, not present in the non-helical 
%case. \note{I dont understand}

The transfer of magnetic energy to magnetic energy ${\mathcal T}_{bb}(Q,K)$ 
seems to follow the same behavior as the velocity field transfer. The 
results show a direct cascade with local transfer of energy from large 
scales to small scales. We note that for the helical case the transfer 
of magnetic energy is larger than the transfer of kinetic energy. 
The likely reason for this behavior is that in the ABC flow the magnetic energy 
at large scales and intermediate scales saturates at higher values than 
in the TG flow, due to the presence of helicity or the dynamo 
$\alpha$-effect. This process will be discussed in more detail 
in Paper II.

\begin{figure}
\includegraphics[width=8cm]{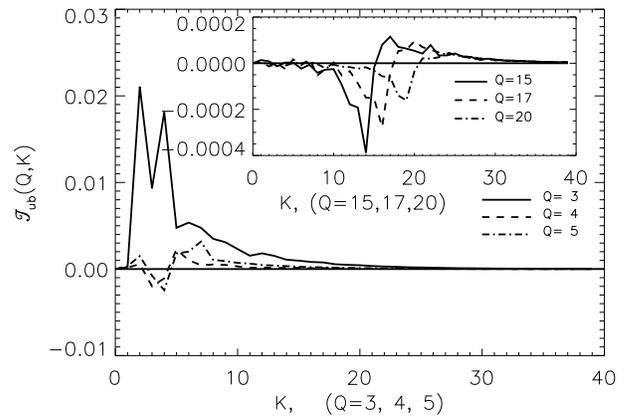}
\caption{\label{fig_06} The transfer of kinetic energy to magnetic energy
${\mathcal T}_{uu}(Q,K)$
for the Taylor Green run.The figure shows the rate that kinetic energy is
transfered from the modes $Q=3,4,5$ (inset modes $Q=15,17,20$)  
to magnetic energy in the modes $K$. }
\end{figure}

\begin{figure}
\includegraphics[width=8cm]{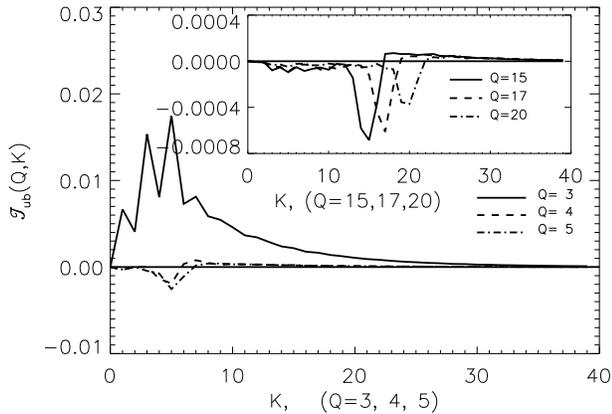}
\caption{\label{fig_07} The transfer of kinetic energy to magnetic energy
${\mathcal T}_{ub}(Q,K)$ for the ABC run. 
The figure shows the rate that kinetic energy is
transfered from the modes $Q=3,4,5$ (inset modes $Q=15,17,20$)  
to magnetic energy in the modes $K$.}
\end{figure}

Next we investigate the transfer of energy from one field to the other, 
by examining the terms ${\mathcal T}_{ub}$ and ${\mathcal T}_{bu}$. 
Because of the anti-symmetric property 
${\mathcal T}_{ub}(Q,K)=-{\mathcal T}_{bu}(K,Q)$, it is sufficient to 
just study the transfer of energy from the velocity field to the magnetic 
field. However, we need to remark that 
unlike the ${\mathcal T}_{uu}(Q,K),{\mathcal T}_{bb}(Q,K)$ 
terms that their dependence on $K$ and $Q$ is the same up to a minus sign,
the behavior of 
${\mathcal T}_{ub}(Q,K)$ as we vary $K$ is not the same as if we vary 
$Q$. Therefore the two behaviors need to be studied separately (i.e., 
the transfer of energy from a velocity mode to two different magnetic 
modes is different from the transfer of energy from 
two different velocity modes to a magnetic mode).
In figure \ref{fig_06} (TG), and \ref{fig_07} (ABC), we show the 
transfer of kinetic energy from the velocity modes $Q=3,4,5,15,17$, and 20 
to all the examined magnetic modes $K$.
%
%%%% FIGURE 6 %%%%%
%%%%%%%%%%%%%%%%%%%
%%%% FIGURE 7 %%%%%
%%%%%%%%%%%%%%%%%%%

A few things should be noted. First, in both runs (ABC and TG) the 
modes associated with the large scale flow ($Q=3$) seem to play a 
dominant role in the transfer of energy from the velocity field to 
the magnetic field. Note also that there is a wider range of magnetic 
field modes into which the forced velocity field modes input energy. 

This is more apparent for the helical flow, that seems better at 
stretching and folding the magnetic field. The cascade in the modes 
inside the inertial range is direct in both cases but with a small difference.
In both cases the large scale velocity field is 
transferring energy to smaller scale magnetic field and receiving energy 
from larger scale magnetic field. However, for the Taylor-Green case
there is very small transfer from one field to the other in the same shell.
On the other hand, in the ABC flow the peak of the transfer from the magnetic 
field to the velocity
field (the negative peaks in figure \ref{fig_07}) is for the same shell. 
Note also that for the $K$-shells 
larger than $Q$, the transfer for all $Q$ follows the same curve. 
This implies that all the small scale velocity modes give energy 
to the magnetic field modes at the same rate. 
This is clearer when we examine the dependence with $Q$.

%%%% FIGURE 8 %%%%%
\begin{figure}
\includegraphics[width=8cm]{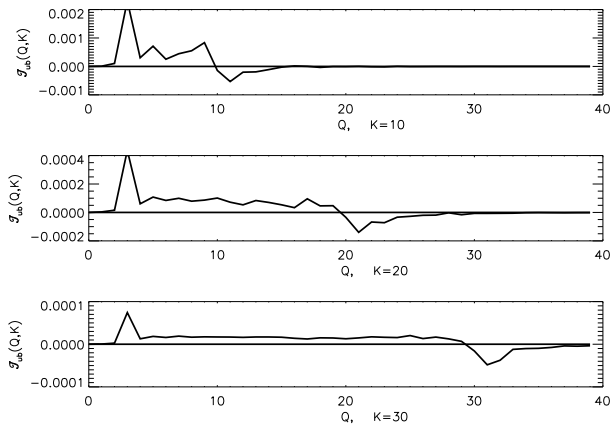}
\caption{\label{fig_08}The transfer of kinetic energy to magnetic energy
${\mathcal T}_{ub}(Q,K,)$ for the Taylor Green run.
The figure shows the rate that kinetic energy is
transfered from the modes $Q$ (x-axis)
to magnetic energy in the modes $Q=10$ (top panel), $Q=20$ (middle panel), 
$Q=30$ (bottom panel).} 
\end{figure}
%%%%%%%%%%%%%%%%%%%
%%%% FIGURE 9 %%%%%
\begin{figure}
\includegraphics[width=8cm]{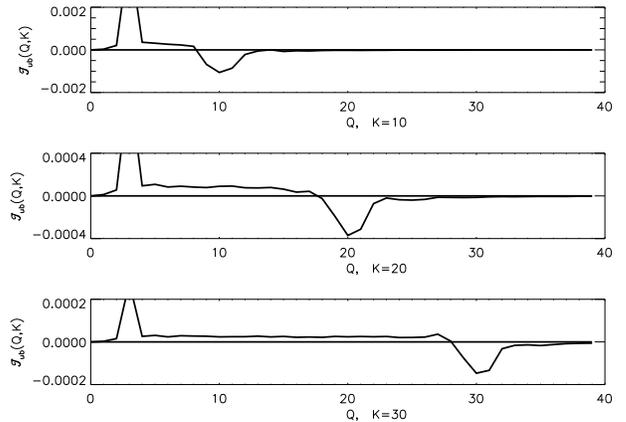}
\caption{\label{fig_09} The transfer of kinetic energy to magnetic energy
${\mathcal T}_{ub}(Q,K,)$ for the ABC run.
The figure shows the rate that kinetic energy is
transfered from the modes $Q$ (x-axis)
to magnetic energy in the modes $Q=10$ (top panel), $Q=20$ (middle panel), 
$Q=30$ (bottom panel).}
\end{figure}
%%%%%%%%%%%%%%%%%%%
%
In figures \ref{fig_08} and \ref{fig_09} we show the same transfer function 
${\mathcal T}_{ub}(Q,K)$ for three values of $K=10,20,30$. 
The energy cascade is also direct (energy going from large scales to small scales),
however it is clear 
from these figures that the transfer from the velocity field to the 
magnetic field is a highly non-local process. Each magnetic field mode $Q$
is receiving energy (positive ${\mathcal T}_{ub}$) from all the velocity 
modes with wave number $K$ smaller than $Q$, with the same rate! 
The only exception is the mechanically sustained large scale velocity field that gives 
even more energy (observe the peak at $k=3$). In fact, 
most of the energy that is transfered from the velocity field
to the magnetic field 
originates from the velocity field modes at $Q=3$ (around $60\%$ for the 
TG run and $75\%$ for the ABC run.)
This energy turns into magnetic energy at several wavenumbers $K$ 
which locally cascades to smaller scales through the ${\mathcal T}_{bb}$ 
term. This bigger contribution of the large scale flow to 
${\mathcal T}_{ub}$ (compared with the contribution of the turbulent 
components) is in good agreement with the suppression of small scale 
velocity fluctuations by the large scale magnetic field, as observed 
in \cite{MininniApJ}. 
However, we need to note that as the scale of the magnetic field becomes smaller
there is more energy input from the turbulent components of the 
velocity field than from the large scale (forced) flow. This 
just follows from the fact that for $K$ large enough, the area below 
the curve with constant ${\mathcal T}_{ub}$ is larger than the peak 
at $Q=3$. It is possible therefore that in the limit of large
inertial range the effect of the forced velocity scales in the
small magnetic scales will not be as strong. 
Finally we note that this mechanism described above 
is different in a kinematic dynamo regime, as 
is shown in Paper II.

In summary, the existence of the long plateau with constant 
${\mathcal T}_{ub}(Q,K)$ at each fixed value of $K$, and the fact 
that all the magnetic wavenumbers $K$ receive energy from the large
scale flow at $Q=3$ points that interactions between the velocity
field and the magnetic field are non-local in Fourier space. 

This non-local behavior of energy transfer from the velocity
field to the magnetic field seems to be absent from the decaying
MHD turbulence case studied by \cite{Debliquy}. In that case although
the ${\mathcal T}_{ub}$ and ${\mathcal T}_{bu}$ were more non-local
than the ${\mathcal T}_{bb}$ and ${\mathcal T}_{uu}$ terms
(since energy was transfered from the former ones in a wider range of shells 
than the later ones), eventually at large separation of wave numbers
the transfer goes to zero. This is very different from the 
plateau behavior we observe in the forced turbulence runs.
We suspect that this difference is due to the fact
that in the mechanically forced turbulence there is a net flux of
energy from the velocity field to the magnetic field that is responsible
for the formation of the plateau which does not exist in the decaying 
turbulence case.

\subsection{A comparison between the transfers}

\begin{figure}
\includegraphics[width=8cm]{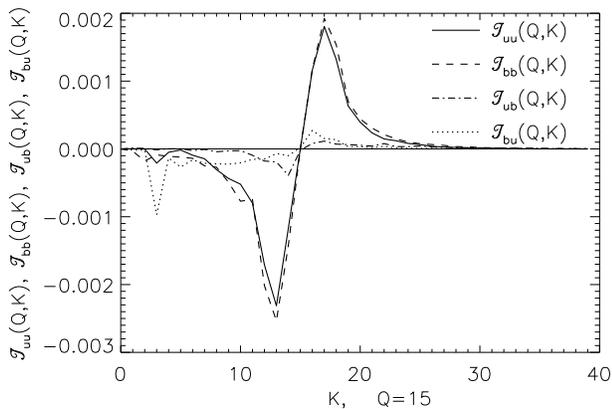}
\caption{\label{fig_10} A comparison of the transfers
${\mathcal T}_{uu}(Q,K),{\mathcal T}_{bb},(Q,K){\mathcal T}_{ub}(Q,K)$ and ${\mathcal T}_{bu}(Q,K)$
for $Q=15$ for the Taylor Green flow.}
\end{figure}

\begin{figure}
\includegraphics[width=8cm]{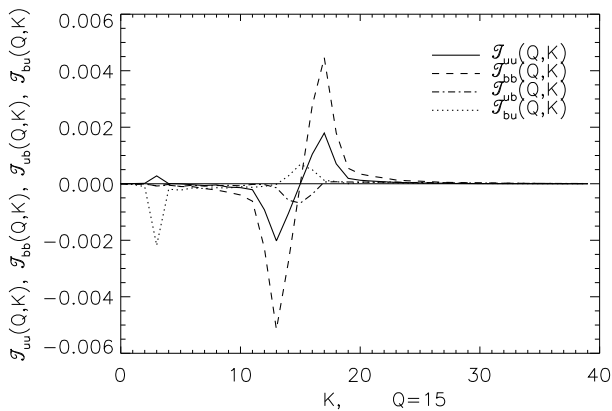}
\caption{\label{fig_11} A comparison of the transfers
${\mathcal T}_{uu}(Q,K),{\mathcal T}_{bb},(Q,K),{\mathcal T}_{ub}(Q,K)$ and ${\mathcal T}_{bu}(Q,K)$
for $Q=15$ for the ABC flow.}
\end{figure}

In the previous section we showed that the transfer of energy from 
the velocity field to velocity field and from the magnetic field to 
magnetic field exhibit a local behavior similar to the transfer
in hydrodynamic turbulence, and the transfer from one field to the other
is exhibiting a non-local behavior. 
In order to draw conclusions we need to compare the magnitude of these transfers. 
Figures \ref{fig_10} and \ref{fig_11} show a comparison of the transfers
${\mathcal T}_{uu}(Q,K),{\mathcal T}_{bb}(Q,K),{\mathcal T}_{ub}(Q,K)$ and
${\mathcal T}_{bu}(Q,K)$ with $Q=15$ for the TG and ABC runs respectively.
%%%% FIGURE 10 %%%%
%%%%%%%%%%%%%%%%%%%
%%%% FIGURE 11 %%%%
%%%%%%%%%%%%%%%%%%%
The local transfers $u$ to $u$ and $b$ to $b$ appear to be of larger 
magnitude than the non-local transfers $u$ to $b$ and $b$ to $u$.
In the case of the ABC flow the magnitude of the $b$ to $b$ transfer 
seems to be twice the magnitude of the $u$ to $u$ transfer. This is 
due to the fact that the magnetic energy in this run is larger than 
in the TG run at large and intermediate scales.

\begin{figure}
\includegraphics[width=8cm]{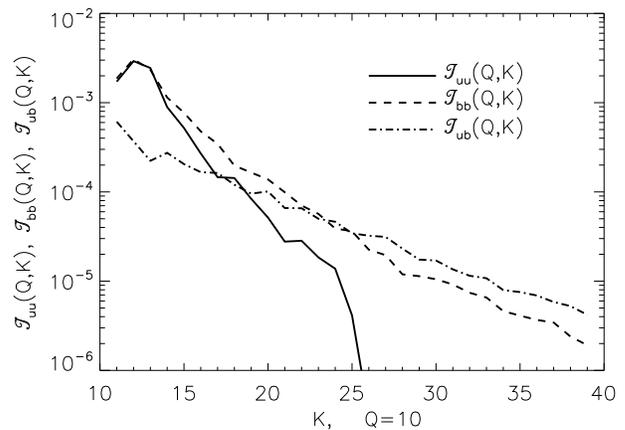}
\caption{\label{fig_12} A comparison of the large $K$ tails  
of transfers ${\mathcal T}_{uu}(Q,K),{\mathcal T}_{bb}(Q,K)$ and 
${\mathcal T}_{ub}(Q,K)$ in a log-linear plot
for $Q=10$ for the Taylor Green flow flow.}
\end{figure}

Figure \ref{fig_12} illustrates the transfer functions 
${\mathcal T}_{uu}(Q,K),{\mathcal T}_{bb},(Q,K)$ and ${\mathcal T}_{ub},(Q,K)$ 
as in  figure \ref{fig_10} (TG flow), but we focus here on the large $K$ 
tail of the transfer and we consider $Q$=10. 
%The fastest drop is for the transfer
%${\mathcal T}_{bu}(Q,K)$ which is the one exhibiting constant
%transfer with respect to $K$ for $K<Q$.
The fastest drop is for the transfer
${\mathcal T}_{uu}(Q,K)$ making it the most 'local' one, next come the
${\mathcal T}_{bb}(Q,K)$ transfer, and finally ${\mathcal T}_{ub}(Q,K)$ 
has the slowest drop. The same result was obtained for the ABC flow 
(not shown here).

\begin{figure}
\includegraphics[width=8cm]{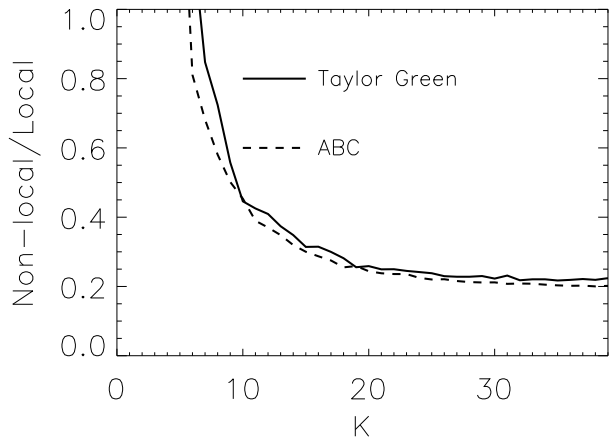}
\caption{\label{fig_13} The ratio $NL/L$ 
of energy received through the non-local transfer ${\mathcal T}_{ub}$ 
to local ${\mathcal T}_{bb}$, for the ABC and TG simulations. 
The small scales receive 20\% of their energy through the non-local 
transfer ${\mathcal T}_{ub}$.}
\end{figure}

Figures \ref{fig_10}, \ref{fig_11}, and \ref{fig_12} 
(as well as a comparison of the nonlocal transfers shown in figures 
\ref{fig_08} and \ref{fig_09} with the local transfers in figures 
\ref{fig_04} and \ref{fig_05} respectively) show that local interactions 
between the same fields are much stronger than nonlocal interactions 
between different fields.
However, nonlocal interactions spread over several shells, 
and the magnetic field at a given scale $K$ can receive (give) energy 
from (to) several velocity field $Q$ wavenumbers (instead of mostly 
the nearest neighbors as is the case for local interactions). 
Figure \ref{fig_13} shows the ratio
\begin{equation}
\frac{NL}{L}(K) = \sum_{Q=1}^{K}{{\mathcal T}_{ub}(Q,K)} \bigg/ 
\sum_{Q=1}^{K}{{\mathcal T}_{bb}(Q,K)} .
\end{equation}
This is the ratio of the total energy that the magnetic field at the shell $K$ receives 
from the velocity field 
through non-local transfer, to the total magnetic
energy received at the same scale through the local direct cascade of 
(magnetic) energy. 
Although in individual shells the local interactions are one 
order of magnitude larger than the non-local transfer, the net amount 
of energy received at a given scale $K$ by the two processes is 
comparable (this ratio is different in a kinematic dynamo regime, as 
will be shown in Paper II). At small scales, the ratio seems to 
settle to a value close to 0.2, indicating that 20\% of the energy 
received by these scales is through the non-local transfer 
${\mathcal T}_{ub}$.
%%%%%%%%%%%%%%%%%%%%%%%%%%%%%%%%%%%%%%%%

\section{ \label{Concs} Conclusions }

\begin{figure}
\includegraphics[width=8cm]{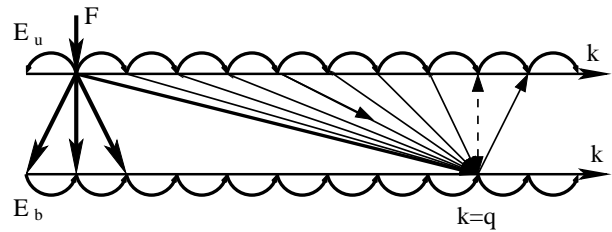}
\caption{\label{fig_14} A sketch of the energy transfer between different
scales and different fields. The thickness of the lines is an indication of the
magnitude of the transfers.
The figure illustrates how energy is transfered
to magnetic modes with wavenumber $k=q$ in the inertial range.
The transfers between same fields is always local and direct.
Each magnetic mode receives energy from all larger in scale velocity modes
and gives to slightly smaller in scale velocity modes. }
\end{figure}

In this paper we examined the transfer of energy in forced MHD turbulence 
between the different scales and fields involved using the results from
numerical simulations in a turbulent steady state sustained by a 
mechanical external force. 
No qualitative differences in the transfer of kinetic to kinetic
(or magnetic to magnetic) energy has been observed, when compared against
the transfer of energy in a hydrodynamic simulation.
These transfers were found to be
always local and direct. However, all kinetic 
energy modes have been observed to give energy to magnetic modes non-locally, 
in the sense that a small scale magnetic field receives 
the same amount of energy from all larger scales of the velocity 
field in the inertial range.
Also each magnetic mode was found to receive a significant
amount of energy from the large scale flow at $|k_F| \sim 3$ 
(the scale of the forcing), an effect that seems to become smaller as we move
to smaller scales in the inertial range. We note that it is the 
non-local interactions that actually sustain the magnetic
field against Ohmic dissipation. 
A summary of our results is sketched in figure \ref{fig_14}.

We have already noted that a different behavior for the
non-local transfers ${\mathcal T}_{ub}$ and ${\mathcal T}_{bu}$ 
was obtained for the mechanically forced turbulence investigated 
in this work, when compared with the decaying turbulence case 
studied by \cite{Debliquy}. Compared with incompressible hydrodynamic 
turbulence, involving only one field and one transfer function, MHD 
turbulence is richer and more complex. It involves two interacting fields, 
several transfer functions, and as a result the energy injected at large 
scales can travel to small scales through several channels. Also the number 
of quadratic ideal invariants is larger, and inverse cascades 
(not present in three dimensional hydrodynamics) can take place. 
This suggests that in MHD flows the particular way the system is 
set-up (e.g. mechanically or magnetically forced, free decaying cases 
without external forces), or even the scale at which the energy is 
injected (compared with the length of the box), might have a direct 
effect in the evolution of the flow and lead to different transfers.

We would also like to comment on the implications of our results to 
the different models of magneto-hydrodynamic turbulence. In the 
present phenomenological models of MHD-turbulence 
\cite{Iroshnikov,Kraichnan,Goldreich,Matthaeus,Boldyrev}, 
locality of the energy transfer is assumed.
That is to say, these models are derived assuming that 
scales of different magnitude do not strongly interact. While this 
assumption seems to be valid for HD-turbulence, this is not necessarily 
true for MHD. As we have shown non-local interactions are present in MHD 
turbulence and control the $u$ to $b$ transfers of energy.
%%%% FIGURE 16%%%%%%%%%
\begin{figure}
\includegraphics[width=8cm]{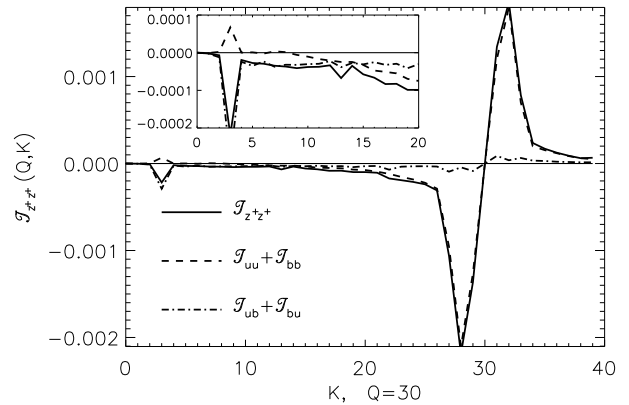}
\caption{\label{fig_15} A comparison of the Els\"asser energy transfer
${\mathcal T}_{z^+z^+}(Q,K)$
with the transfers from the local 
${\mathcal T}_{uu}+{\mathcal T}_{bb}$
and the non-local 
${\mathcal T}_{ub}+{\mathcal T}_{bu}$
contributions.
The inset is showing a blow up of the small $K$ tail where 
the non-local interactions are more dominant.} 
\end{figure}
%%%%%%%%%%%%%%%%%%%%%%
However, these non-local interactions are smaller in amplitude
and most of the input of energy to the magnetic field comes from 
the large scale flow and then cascades to smaller scales making the 
assumption of locality justified to some extend. 
However non-local $u-to-b$ need to be considered to have a proper description
of the energy cascade. 
To illustrate this we show in figure \ref{fig_15} the energy transfer 
in terms of the Els\"asser variables ${\bf z}^\pm={\bf u}\pm {\bf b}$, often used in 
turbulence models and we compare it with the contributions from 
$u$ to $u$, $b$ to $b$, $b$ to $u$, and $u$ to $b$. In the figure we plot 
${\mathcal T}_{z^+z^+}\equiv -\int {\bf z^+}_K{\bf z^-\nabla z^+}_Q  d{\bf x}$ 
and compare it with the energy transfer due to the local transfer terms 
${\mathcal T}_{uu}+{\mathcal T}_{bb}$ and the non-local transfer terms 
${\mathcal T}_{ub}+{\mathcal T}_{bu}$. The local transfer terms appear 
to be dominant, except in the tails where the transfer of Els\"asser 
variables is dominated by the non-local transfers between the magnetic 
and kinetic energies.
This tails, although with small amplitude, cannot 
be completely neglected, as shown by the $NL/L$ ratio of figure 13. The non-local tail 
in the transfer gives a net contribution of energy at magnetic small 
scales of roughly 1/5 when compared with the local transfer.

Finally, we would like to say that our results were based on numerical
simulations of moderate Reynolds number much smaller than what is
observed in most physical phenomena. We already noted that due to the small
inertial range we cannot test self similarity that would require to 
compare the transfers (i.e. ${\mathcal T}_{uu}(Q,K_1)$, 
${\mathcal T}_{uu}(Q,K_2)$ )
to wave numbers that are both significantly away from each other 
($K_1 \ll K_2$) 
and away from the forced  and dissipative scales ($K_F \ll K_1$ and 
$K_2 \ll k_{\eta}$).

Finally the transfer of 
magnetic helicity and cross-helicity and their effect on the
turbulence dynamics is also worth studying, but we leave these issues
however for our future work.

%However, we expect that the qualitative features of our results should
%hold in higher Reynolds numbers $Re,R_M$. 

\begin{acknowledgments}
The authors are grateful to J. Herring for valuable discussions
and his careful reading of this document.
Computer time was provided by NCAR. The NSF grant CMG-0327888
at NCAR supported this work in part and is gratefully acknowledged.
\end{acknowledgments}

\bibliography{ms}% Produces the bibliography via BibTeX.

\end{document}